\documentclass[pr,twocolumn]{revtex4}
\usepackage{bm}
\usepackage{graphicx}
\usepackage{amssymb}
\usepackage{amsmath}
\usepackage{eufrak}
\usepackage{color}
\usepackage[utf8]{inputenc} %german �äö...
\usepackage{hyperref}
\usepackage{pifont}
\usepackage{ulem}
\usepackage{epstopdf}

\newcommand{\nix}[1]{}

\newcommand{\MD}[1]{{\color{red} #1}}

\newcommand{\kp}{$\bm{k} \cdot \bm{p}$}

\begin{document}

%{\color{red}
%{\color{blue}

\title{Determination of hole $g$-factor in InAs/InGaAs/InAlAs quantum wells by magneto-photoluminescence studies \\
\MD{}}

\author{Ya.\,V.~Terent'ev$^{1,2}$, S.\,N.~Danilov$^1$,  M.\,V.~Durnev$^2$, J.~Loher$^1$,
D.~Schuh$^1$, D.~Bougeard$^1$,
 S.\,V.~Ivanov$^2$, and
S.\,D.~Ganichev$^1$}
\affiliation{$^1$ Physics Department, University of Regensburg,
93040 Regensburg, Germany}
\affiliation{$^2$Ioffe Institute,
%Russian Academy of Sciences,
194021 St.~Petersburg, Russia}

\begin{abstract}
Circularly-polarized magneto-photoluminescence (magneto-PL) technique has been applied to
investigate  Zeeman effect in InAs/InGaAs/InAlAs  quantum wells (QWs)  in Faraday geometry.
Structures with different thickness of the QW barriers have
been studied in magnetic field parallel and tilted with respect to the sample normal. 
Effective electron-hole  $g$-factor has been found by measurement of
splitting of polarized magneto-PL lines. Land\'e factors of electrons have been calculated using  the 14-band \kp~method and
$g$-factor of holes was determined by subtracting the calculated contribution
of the electrons from the effective electron-hole $g$-factor.
Anisotropy of the hole $g$-factor has been studied applying tilted magnetic field.

\end{abstract}
\pacs{xxx}

\maketitle{}

\section{Introduction}

Heterostructures based on InAs possess series of unique properties caused by a narrow bandgap. These properties include
high carrier mobility and a strong spin-orbit interaction making the system a promising candidate for high frequency
electronics, optoelectronics and spintronics application. One of the most interesting objects in this area are  
type-I quantum 
wells (QWs) based on a InAs/InGaAs/InAlAs two-step bandgap engineering, where In content can be varied from
30 to 80\%~\cite{Inoue1991}. Such structures exhibit bright photoluminiscence
in mid-infrared range~\cite{Tournie1992,Terent'ev2014} demonstrate high-mobility two-dimensional electron
gas~\cite{Richter2000,Heyn2003,Moller2003,Hirmer2011}, pronounced spin-dependent
optical~\cite{Tsumura2006,Terent'ev2014,Terent'ev2015} 
and transport~\cite{Hu2003,Nitta2003,Wurstbauer2009,Wurstbauer2010,Ganichev2009,Olbrich2012} phenomena.
Determination of Land\'e factors of both types of the carriers is the cornerstone for the studies of spin-related 
phenomena. As for InAs QWs, electron $g$-factor in this type of heterostructures is well-studied by different techniques.
To date, reported values of electron $g$-factor obtained by magneto-transport and terahertz experiments
range from $g_{e}$= -3 to $g_{e}$= -9 depending on In content in QW barrier~\cite{Moller2003,Wurstbauer2009,Nitta2003}. 
Moreover, experimentally obtained  Land\'e factors are consistent with the values calculated in the framework of \kp~method.

In contrast to electrons, determination of the hole $g$-factor $g_h$ is still a challenging task.
 There are no available experimental data as well as reliable theoretical calculations.
The picture becomes even more intriguing in light of the previous magneto-optical experiments ~\cite{Tsumura2006,Terent'ev2014}.
They indicate surprisingly small magnitude of the effective electron-hole $g$-factor which is the difference between  $g_{h}$ and
 $g_{e}$.

Here we report on  studies of 
InAs/In$_{0.75}$Ga$_{0.25}$As/In$_{0.75}$Al$_{0.25}$As QW structures  by polarization-resolved magneto-PL,
which enables direct measurement of effective electron-hole $g$-factor. We have determined Land\'e factor of holes
combining obtained experimental data with theoretical calculations of an electron contribution. 
We have obtained the dependence of electron and hole $g$-factors on QW barrier width.
We have investigated anisotropy of the hole $g$-factor in tilted magnetic field, and shown that the values of $g_h$ in tilted magnetic field are in agreement with prediction of close-to-zero hole Zeeman splitting in magnetic field lying in QW plane.

\section{Samples and Experimental technique}

\begin{figure}[b]
\includegraphics[width=0.9\linewidth]{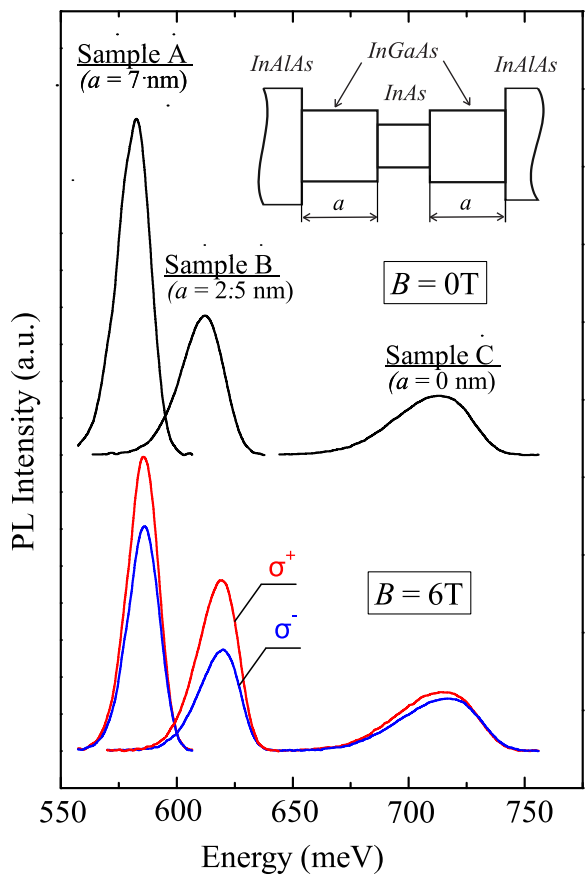}
\caption{Photoluminescence spectra and circular-polarized magneto-PL spectra of samples A, B and C
with different width of InGaAs barriers, measured at magnetic field $B = 6$~T perpendicular to QW plane. The inset 
shows the band diagram of the samples active region.}
\label{fig01}
\end{figure}

Experimental samples were fabricated by molecular beam epitaxy onto a fully
relaxed In$_x$Al$_{1-x}$As/(001)GaAs graded
buffer~\cite{Capotondi2005,Terent'ev2014}
with a stepwise increase of the In content ($x$ = 0.05 to $x$ =
0.75) over 1~$\mu$m. {The structure of QW is} sketched in
the inset of Fig.~\ref{fig01}.
An  In$_{0.75}$Ga$_{0.25}$As quantum size part embedded in between  In$_{0.75}$Al$_{0.25}$As
layers features a symmetrically inserted and compressively strained InAs QW of 4 nm. 
%Such an
%approach on structure design enables the fabrication of high quality defect-free and strain relaxed virtual substrates of high indium
% content that allow an effective collection of photogenerated carriers into high quality QW.
 A set of samples with the different thickness of 
 In$_{0.75}$Ga$_{0.25}$As barrier $a$ was grown, where $a$ is set to 7, 2.5 and 0 nm. The corresponding
structures are labeled A, B and C.

PL was excited by emission of a laser diode operating in the
\textit{cw} mode at wavelength $\lambda = 809$~nm and detected with a Fourier Transform Infrared
(FTIR) spectrometer.  The laser beam was focused to a 1-mm diameter spot on the sample. The
excitation intensity $W_{exc}$ was 100~mW. An external magnetic field up to 6~T was applied perpendicularly
to the wafer or was inclined at an angle of 40$^{\circ}$ to the direction of sample growth. PL emission 
having wave vector directed along magnetic field was detected (Faraday geometry).
The sample temperature was kept as low as 2~K. Right- and left-handed circular polarized emission spectra
were recorded applying a quarter wave ZnSe Fresnel
rhomb~\cite{Ganichev1,Ganichev2}. 

\section{Experimental results}

Bright  PL was detected from all the  samples.
Its contour is close to Gaussian function and
slightly asymmetric being broadened at low energy slope, see Fig.~\ref{fig01}. PL peak energy takes the value of 
 582.6 meV, 616.5 meV and 713.6 meV in structures  A, B and C, respectively. So PL peak energy increases with decrease of
InGaAs barrier width. Emission intensity
also varies in samples with different  barrier width and decays with its reduction.
Both trends correlate with increase of PL peak  full width at half maximum (FWHM)  that takes
the value of  18, 22 and 40 meV,  in samples A, B and C, respectively. 
Application of external magnetic field $\bm B$  in direction perpendicular to QW plane results in substantial changes in the PL spectrum.
It experiences magnetic-field-induced splitting into circular polarized components, which is different in structures A, B and C,  see Fig.~\ref{fig02}.
Similar to PL peak energy and FWHM, the splitting depends on InGaAs barrier width.
It is extremely small in structure A with the largest InGaAs barrier but is well-pronounced in sample C, where this layer is absent, taking intermediate
value in structure B. Interestingly, the splitting is a non-linear function of the magnetic field and its dependence on
magnetic field is different in all three samples, see Fig.~\ref{fig02}.
Note that minimal circular polarization of PL emission was detected in structure C while it 
possesses the largest splitting. The inclination of the magnetic field used for the analisis of  $g_{h}$ anisotropy
critically diminishes the splitting of a magneto-PL peak in samples B and C, however does not affect
it in sample A, see Fig.~\ref{fig02}.

Besides the splitting of the PL contour into circular-polarized components, emission spectra experience blue shift, which corresponds to the diamagnetic shift of electron and hole energy levels in magnetic field. The shift has a quadratic dependence on the magnetic field,
see  Fig.~\ref{fig03}.

\section{Discussion}

The observed PL peak  originates from direct optical transitions between the ground electron  $e1$ and the heavy
hole $hh1$ subbands, according to calculations of the optical transition 
energy. It is important to note, that in our case optical transitions between free-carrier states dominate,
in contrast to wide band systems, where the exciton recombination prevails~\cite{Terent'ev2014}.

%The broadening of PL peaks is explained mainly in terms of
%broadening of the energy levels due to the variation of the QW width and indium composition as well as scattering
%by charged centers at QW interfaces. Apparently  in the sample C large FWHM of the PL peak  is associated with this 
%scattering mainly that can also be responsible for strong depolarization of the emission.

Observed in experiment quadratic magnetic field dependence of the PL peak diamagnetic shift (Fig.~\ref{fig03}) indicates Coulomb localization
of the photoexcited carriers  ~\cite{Wang1994,Sugawara1997}. In our structures carriers can be trapped to localization centers which  emerge
due to inhomogeneity of the InAs QW or  presence of charged centers at the interfaces of QW.

\begin{figure}[t]
\includegraphics[width=0.8\linewidth]{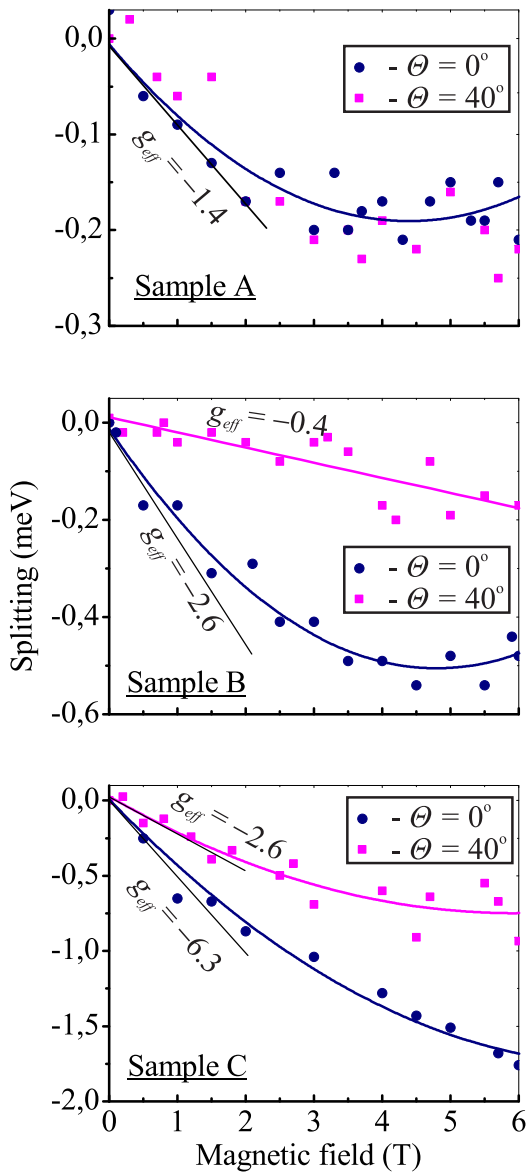}
\caption{
Zeeman splitting of magneto-PL peak for structures A, B and C with different width of InGaAs barriers
measured at magnetic field perpendicular to QW plane and tilted by $\theta = 40^\circ$ with respect to the sample normal. Curves present polynomial fit to the experimental points.
Straight lines show linear fit at small magnetic fields, which gives effective $g$-factors of free carriers.
}
\label{fig02}
\end{figure}

\begin{figure}[t]
\includegraphics[width=0.8\linewidth]{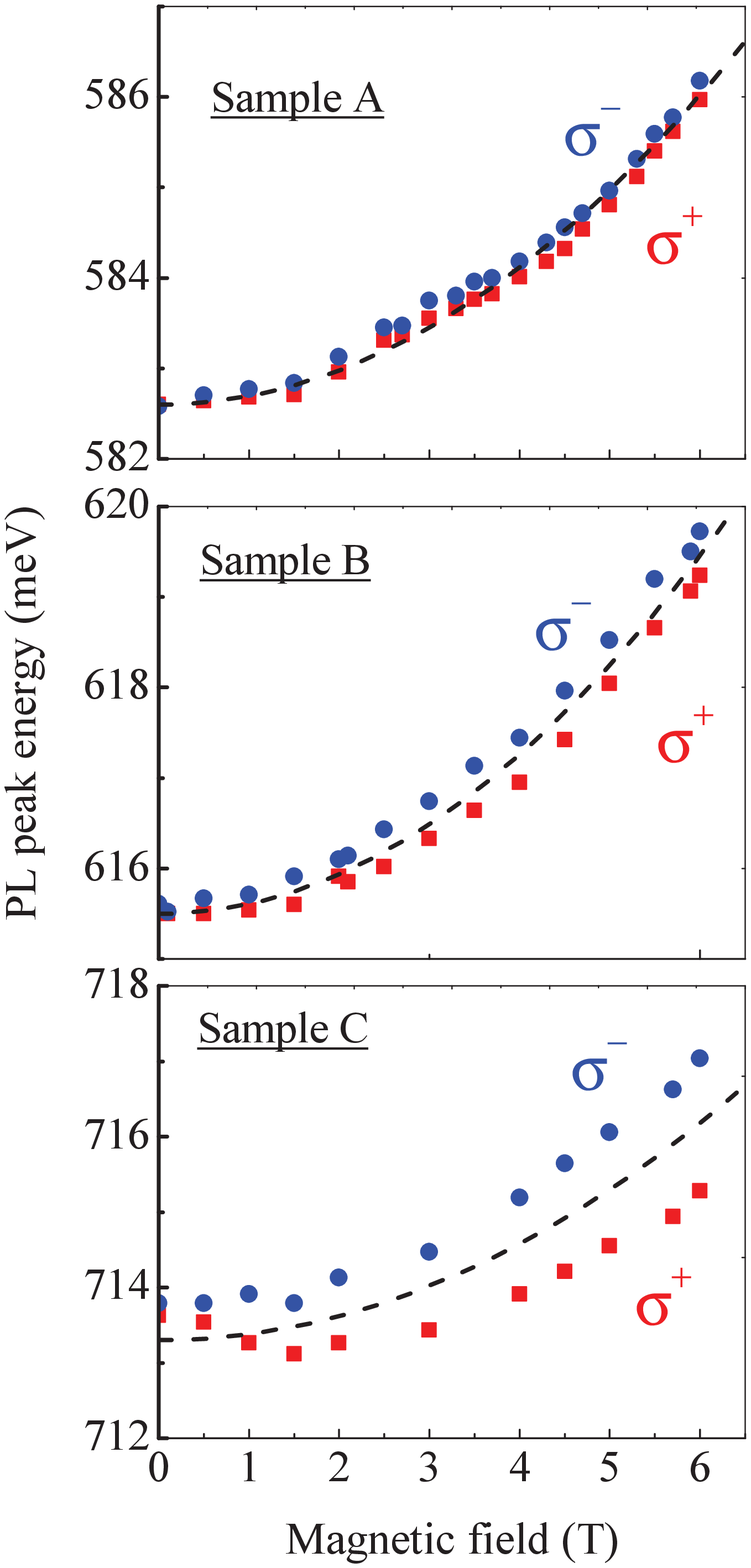}
\caption{
Circular polarized PL peaks energies as a function of magnetic field. Curves present parabolic fitting $E = \alpha_{\rm dia} B^2$ of the diamagnetic shift of PL lines, which is determined as a half-sum of the $\sigma^+$ and $\sigma^-$ components. The values of $\alpha_{\rm dia}$ are listed in Tab.~\ref{tab1}.
}
\label{fig03}
\end{figure}

\begin{figure}[t]
\includegraphics[width=0.9\linewidth]{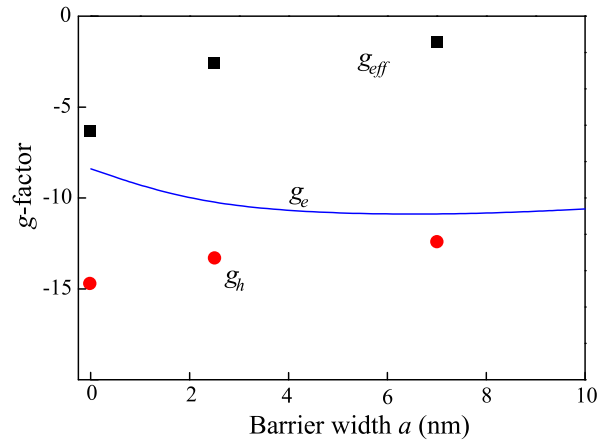}
\caption{$g$-factors in samples under study:
squares {represent the experimentally measured effective electron-hole $g$-factor $g_{\rm eff}$, the solid line stands for the theoretically calculated electron $g$-factor, and circles show the heavy-hole $g$-factor derived using formula $g_h =  g_{\rm eff} + g_e$.}
}
\label{fig04}
\end{figure}

Detected splitting of PL lines in two circular polarizations reflects spin splitting of conduction 
and valence bands. The linear region of the Zeeman splitting (see Fig.~\ref{fig02}) gives the value of the effective electron-hole $g$-factor ($g_{\rm eff}$), which is equal to the difference of $g$-factors of the carriers that take part in the optical recombination, $g_{\rm eff} = g_h - g_e$~\cite{Terent'ev2014}. The extracted values of $g_{\rm eff}$ as a function of an InGaAs barrier width are presented in Fig.~\ref{fig04} and Tab.~\ref{tab1}.
It is clearly seen that the absolute value of $g_{\rm eff}$ tends to increase with decreasing width of InGaAs barrier. 

While observed nonlinear character of Zeeman splitting in magnetic field {$B \gtrsim 2$} is out of scope of present paper, it is worth mentioning  that highly nonlinear Zeeman splitting of excitons was also detected in AlGaAs/GaAs and InGaAs/GaAs systems and the model was suggested that is based on a spin-dependent field-induced admixture between the light- and heavy-hole valence bands ~\cite{Traynor1997}.

Large FWHM of the PL peak and its decay detected in sample C is explained in terms of  effective scattering by charged centers at QW interfaces.
In this structure QW interfaces are formed by InAs and InAlAs layers having large lattice mismatch and therefore are characterized by large
defects density.  Apparently scattering on these defects is responsible for the  strong decay of optical recombination efficiency, increase of FWHM
as well as depolarization of the emission in the case of the magnetic field applied.

\begin{table}[htb]
\caption{\label{tab1} The values of electron and hole $g$-factors ($g_{\rm eff}$, $g_e$, $g_h$) and diamagnetic shifts ($\alpha_{\rm dia}$) extracted from the analysis of experiment. The values $g_e^*$ and $g_h^*$ are derived using Eq.~\eqref{eq:g_tilt} from the analysis of Zeeman splitting in tilted magnetic field.
}
\begin{ruledtabular}
\begin{tabular}{cccc}
  & Sample A & Sample B & Sample C \\
$g_{\rm eff}$ & -1.4  & -2.6 & -6.3 \\
$g_{e}$ & -11  & -10.7 & -8.4 \\
$g_{h}$ & -12.4  & -13.3 & -14.7 \\
$g_{e}^*$ & --  & -9 & -13.2 \\
$g_{h}^*$ & --  & -9.4 & -15.8 \\
$\alpha_{\rm dia}$ (meV/T$^2$) & 0.095 & 0.11 & 0.08
 \end{tabular}
\end{ruledtabular}
\end{table}

Now we turn to separate determination of electron and hole $g$-factors. {It was found that the $g$-factor of hole is extremely sensitive to the separation between heavy-hole and light-hole quantization levels, which in turn depends on the unknown strength of strain fields and localization potential. Hence we only calculate $g_e$, which is less affected by the localization potential and strain, and therefore can be evaluated with much higher accuracy. Then
we estimate the value of $g_{h}$ using experimentally determined $g_{\rm eff}$ and 
\begin{equation}
\label{eq:gh}
 g_{\rm eff} = g_h - g_e\:.
\end{equation}
%Remarkably, we can evaluate $g_h$ using experimentally obtained values of $g_{eff}$ if we know $g_e$. 
%The later parameter can be calculated with high accuracy, in contrast to $g_h$.
Computation method is based on numeric diagonalization of the 14-band \kp~Hamiltonian in the presence of magnetic field~\cite{DurnevPRB2014, Durnev2014}. We use the developed in Ref.~\cite{DurnevPRB2014} 14-band \kp-model for calculation of electron and hole states at zero magnetic field, and use obtained wave functions to calculate the Zeeman splitting at small magnetic fields in the framework of perturbation theory (the approach is analogous to the one used to calculate heavy-hole and light-hole $g$-factors in the framework of Luttinger Hamiltonian, see Eqs.~(9a), (9b) of Ref.~[\onlinecite{Durnev2014}]). The band parameters of InAs and its alloys were taken from Refs.~\cite{Meyer2001, Walle1989}, and confining potentials for electron and holes were calculated using the model of Ref.~\cite{Walle1989} taking into account elastic strain present in the structure. The interband matrix elements of momentum operator were taken from Ref.~\cite{Jancu2005}. }

\begin{figure}[t]
\includegraphics[width=0.9\linewidth]{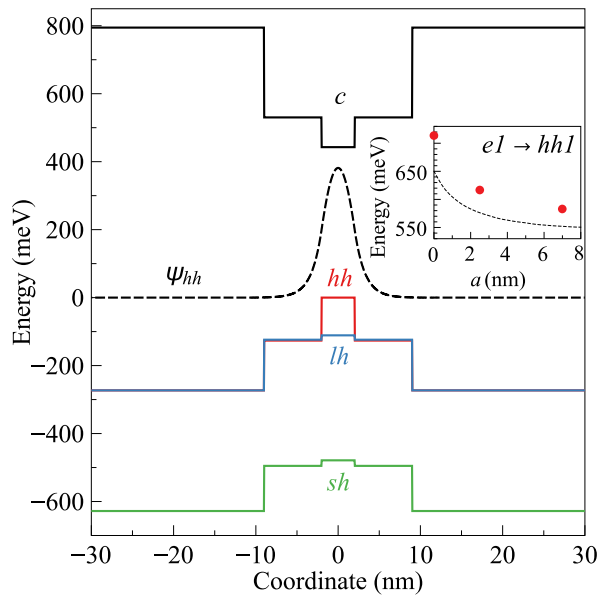}
\caption{Band diagram of the quantum well with $a = 7$~nm. The heavy-hole wave function is shown by dashed line. The inset demonstrates the calculated optical transition energy (dashed line) and the experimental points for three samples under study.  
}
\label{fig:band_diagram}
\end{figure}

The evaluated band diagram, wave function of heavy hole at zero in-plane momentum $\bm k_\parallel = 0$, and optical transitions energies are presented in Fig.~\ref{fig:band_diagram}. Theoretical model gives close-to-experiment values of optical transitions energies and the in-plane electron mass $m_e$ ($m_e = 0.038~m_0$ as determined from transport measurements in similar structures~\cite{Richter2000, Moller2003, Hu2003}). It also provides a good agreement between theoretical and experimentally measured values of $g_e$ ~\cite{Danilov2015, Moller2003, Hu2003}. The discrepancy between theoretically calculated and experimental values of optical transitions energies is possibly attributed to the presence of in-plane localization potentials, which lead to increase of the PL peak energy.

Evaluated values of hole Land\'e factors $g_h$ as well as theoretical values of $g_e$ and experimentally measured $g_{\rm eff}$ are presented in Fig.~\ref{fig04} and Tab.~\ref{tab1}.  
As we mentioned above theoretical calculations do not allow to obtain $g_h$ with sufficient precision. Its values obtained in the framework of 14-band \kp-model, used for evaluation of $g_e$,
 lie in the range of $g_h = - 2 \div - 8$ for the studied structures~\cite{Danilov2015}, and differ significantly from those given in Tab.~\ref{tab1}. 
% For reference, calculated $g_h \approx 0$ for a three-dimensional hole bound to acceptor in an InAs layer~\cite{rodina1998, gelmont1974}.

%

In order to study anisotropy of hole  $g$-factor we have carried out experiments in a tilted magnetic field.
While electron Zeeman splitting is known to be almost independent of the direction of applied magnetic field~\cite{ivch_book}, the spin splitting of a heavy hole bound to QW potential must be sensitive to a normal component of magnetic field only, because the in-plane heavy-hole $g$-factor is close to zero in III-V quantum wells~\cite{Marie01}. Hence tilted magnetic field should result in modification of $g_{\rm eff}$. {Indeed, heavy-hole $g$-factor in a magnetic field tilted by an angle $\theta$ with respect to QW normal is
\begin{equation}
\label{eq:gh_theta}
\tilde{g}_{h} (\theta) = \sqrt{g_{h}^2 \cos^2 \theta + g_{h,\parallel}^2 \sin^2 \theta}\:,
\end{equation}
where $g_{h}$ and $g_{h,\parallel}$ are the components of the $g$-factor tensor for $\bm B$ parallel to the growth axis $z \parallel [001]$ and $\bm B$ oriented in the plane of QW. Since $g_{h,\parallel}$ is close to zero, $\tilde{g}_{h} \approx g_{h} \cos \theta$ and its absolute value must be reduced at $\theta \neq 0$.

For example, the 40$^{\circ}$ tilt of the field is expected to result in reduction of $g_h$ 
from -13.3 and -14.7 to -10.2 and -11.3 for samples B and C, respectively. In turn, assuming $g_e$ is independent of $\theta$, the value of $g_{\rm eff}$ is predicted to change to 0.5 and -2.9 for these samples. Comparison with Fig.~\ref{fig02} shows that these estimated values are close to the values observed in experiment.

Anomalous behaviour of the Zeeman splitting was observed in structure A only. As seen from Fig.~\ref{fig02}, the effective Zeeman splitting is almost unaffected by the tilt of $\bm B$ in this sample. It may indicate, that the heavy hole in this sample is localized by rather a three dimensional potential than the potential of QW. This suggestion is consistent with the value of the in-plane localization length of hole $l_{h} \approx 10$~nm (see below for details), which in the case of sample A is less than the effective localization in the $z$-direction (see Fig.~\ref{fig:band_diagram}). The possible source for such a three-dimensional confinement is a Coulomb potential of charged centers in quantum well layers.

With the use of Eqs.~\eqref{eq:gh} and \eqref{eq:gh_theta} it is possible to evaluate $g_h$ and $g_e$ independently without theoretical calculations by measuring $g_{\rm eff}$ at two different angles $\theta$:
\begin{eqnarray}
\label{eq:g_tilt}
g_h &=& \frac{g_{\rm eff}(\theta_1)-g_{\rm eff}(\theta_2)}{\cos\theta_1- \cos\theta_2}\:, \\
g_e &=& \frac{g_{\rm eff}(\theta_1) \cos\theta_2 - g_{\rm eff}(\theta_2) \cos\theta_1}{\cos\theta_1 - \cos\theta_2} \nonumber \:.
\end{eqnarray}
The calculated values are presented in Tab~\ref{tab1}.

Let us finally analyze the diamagnetic shift of PL lines, see Fig.~\ref{fig03}. The value of diamagnetic shift is given by a half-sum of the $\sigma^+$ and $\sigma^-$-polarized components and is well fitted by the quadratic function $E_{\rm dia} = \alpha_{\rm dia} B^2$. The extracted values of $\alpha_{\rm dia}$ are listed in Tab.~\ref{tab1}. To derive theoretical expression for $\alpha_{\rm dia}$ we will use a simple model of the carriers bound by a parabolic in-plane potential in the form~\cite{Sugawara1997, PhysRevB.57.9088}
\begin{equation}
\label{eq:pot}
V(\bm r_{n}) = \frac{\hbar^2}{2 m_{n} l_n^2} \frac{r_{n}^2}{l_{n}^2}\:,
\end{equation}
where $\bm r_{n}$ is the in-plane coordinate, $m_{n}$ is the effective in-plane mass, $l_{n}$ is the in-plane localization length of an electron ($n = e$) and heavy-hole ($n = h$), and $\hbar$ is the Planck constant. Making the Peirels substitution for the carrier wave vector in magnetic field and solving the Schr\"odinger equation with potential~\eqref{eq:pot} we find for the diamagnetic coefficient $\alpha_{\mathrm{dia}, n}$
\begin{equation}
\alpha_{\mathrm{dia}, n} = \frac{e^2 l_{n}^2}{8 m_n c^2}\:,
\end{equation}
where $e$ is the electron charge, and $c$ is the speed of light. Since in QWs under study $m_e \ll m_{h}$ (electron mass $m_e = 0.038~m_0$~\cite{Richter2000, Moller2003, Hu2003}, heavy-hole mass $m_{h} = 0.085~m_0$ is deduced from the \kp~calculations for sample A) we conclude that the main contribution to the diamagnetic shift results from a confined electron. Taking experimentally measured values of $\alpha_\mathrm{dia}$ we find that the in-plane localization length for an electron is $l_{e} \approx 13$~nm in all studied samples. Localization length for a hole is $l_{h} = (m_e/m_{h})^{1/4} l_{e} \approx 10$~nm and is even smaller than $l_{e}$.

\section{Conclusions}

To summarize, series of magneto-optical experiments have been carried out on narrow gap 
InAlAs/InGaAs/InAs QWs with different width of InGaAs barrier in both perpendicular and tilted magnetic fields. 
Effective electron-hole $g$-factor is measured directly from the splitting of magneto-PL line into
circularly-polarized terms. The values of electron $g$-factor $g_e$ are calculated theoretically while the $g$-factor of
holes is estimated by extracting $g_e$ from the total splitting.
%Thus value of $g$ factors for both types of the free carriers are obtained.
Experiments in tilted magnetic field were used to investigate anisotropy of the heavy hole $g$-factor.

\acknowledgements

The authors are grateful to S. A. Tarasenko for interest in the work and useful discussions.

The authors gratefully acknowledge financial support from DFG SFB 689, Ya.V.T. and S.V.I. are thankful to the support from RSF (Project 15-12-30022). M.V.D. acknowledges financial support from RFBR project No. 16-32-60175 and the Dynasty foundation.

{
%\color{red} Questions to clarify:\\
%$\diamond$ to address in text the $g$-factor difference for different thicknesses?\\
%$\diamond$ to address calibration of $\sigma^+$ and $\sigma^-$?\\
%$\diamond$ to emphasize GaAs substrate (request of Yakov)
%}

\end{document}